\documentclass[aps, prb, reprint, showkeys, nolongbibliography]{revtex4-2}

\pdfoutput=1

\usepackage[utf8]{inputenc}
\usepackage[usenames,dvipsnames]{color}
\usepackage{hyperref}
\usepackage{amssymb}
\usepackage{amsmath}
\usepackage{multirow}
\usepackage{epsfig}
\usepackage{rawfonts}
\usepackage{latexsym}
\usepackage{theorem}
\usepackage{natbib}
\usepackage{graphicx}
\usepackage{float}
\usepackage{mathtools}
\usepackage[normalem]{ulem}
\usepackage[section]{placeins}

\begin{document}

\title{Quartic energy band engineering in artificial semiconductor honeycomb lattices}
\date{\today}

\author{Emre Okcu$^{1}$, Emre Mesudiyeli$^{1}$, Hâldun Sevinçli$^{2}$, A. Devrim Güçlü$^{1}$}
\affiliation{$^{1}$Department of Physics, \.{I}zmir Institute of Technology, 35430 Urla, \.{I}zmir, T\"{u}rkiye \\
$^{2}$Department of Physics, Bilkent University, 06800 Bilkent, Ankara, T\"{u}rkiye}

\begin{abstract}
	Artificially engineered lattices provide a flexible platform for reproducing and extending the electronic behavior of atomic-scale materials. Artificial graphene systems, in particular, mimic graphene-like linear dispersion with tunable Dirac cones and offer a route to realizing more exotic band structures. Here we examine the emergence of quartic energy dispersion in artificial graphene heterostructures using analytical modeling and numerical solutions of the effective Hamiltonian. We identify three distinct quartic band types—Mexican-hat-shaped (MHS), purely quartic, and non-MHS quartic bands—and determine the conditions under which each arises. We find that a staggered honeycomb lattice supports all three classes of quartic dispersion, whereas its planar counterpart yields only purely quartic and non-MHS forms. These results demonstrate the feasibility of engineering quartic band edges in artificial lattices and clarify how lattice geometry can be used to tailor their characteristics.

\end{abstract}


\maketitle
\section{Introduction}

Recent advances in semiconductor-based superlattices have opened new avenues for developing quantum simulators \cite{1, 2, 3, 4, 5, 6, 7, 8, 9, 10, 11}. 
Such quantum simulators provide a highly controllable platform enabling the replication of complex systems that are challenging to fabricate. 
Among these, artificial graphene nanostructures enable the engineering of Dirac cones \cite{12, 13, 14, 15, 16, 17, 18} and investigation of strongly correlated quantum many-particle phases, magnetic phenomena \cite{19,Bulut,gokhan_2026} and finite size effects, by tuning system parameters such as lattice constant, dot radius, and potential depth. 
Artificial graphene structures are typically fabricated by imposing a periodic potential onto a two-dimensional electron gas (2DEG), often using lithographically patterned gate electrodes or nano-patterned substrates. \cite{gibertini2009engineering, 17, wang2018gaas}.

Beyond the massless Dirac dispersion of graphene, hexagonal lattices can host a variety of unconventional band structures.
Electrically biased bilayer graphene provides a realization of MHS band edges through inversion-symmetry breaking and field-induced band rearrangement~\cite{Castro2007,McCann2006}.
Hexagonal structures of group-V elements exhibit MHS quartic valence bands without the need for an external field~\cite{zhu_prl_2014,kamal_prb_2015,ozcelik_prb_2015,akturk_prb_2015,sevincli_2017}.
Hexagonal group-III--VI monolayers and group-IV--V monolayers also display quartic valence bands, some of which are MHS while others are purely quartic~\cite{demirci_prb_2017,ozdamar_prb_2018,damljanovic2025existence}.
MHS quartic dispersions have also been reported for other 2D materials such as $\alpha$-SnO~\cite{Seixas2016}, KTlO~\cite{Song2019KTlO}, GeP$_3$~\cite{Wang2021GeP3}, and penta-graphene~\cite{Zhang2015Penta}.

In all these materials, the quartic dispersion forms around the center of the Brillouin zone and exhibits a strong Van Hove singularity.
Such inverse-square-root singularities are typically associated with strongly correlated many-body effects, including superconductivity \cite{mazziotti2015superconductivity,VHS1,VHS2,VHS3}, spontaneous magnetization \cite{sevincli_2017}, Wigner crystallization \cite{21}, and high thermopower \cite{sevincli_2017,TP}.
The unusual transport and magnetic responses arising from this dispersion—such as enhanced thermopower, ferromagnetic instability, and disorder-driven localization near the band edge—have been analyzed in both model Hamiltonians and first-principles studies \cite{sevincli_2017,Canbolat2022,Polat2023,cao_prl_2015}.

\begin{figure}
	\centering
	\includegraphics[width=1.0\columnwidth]{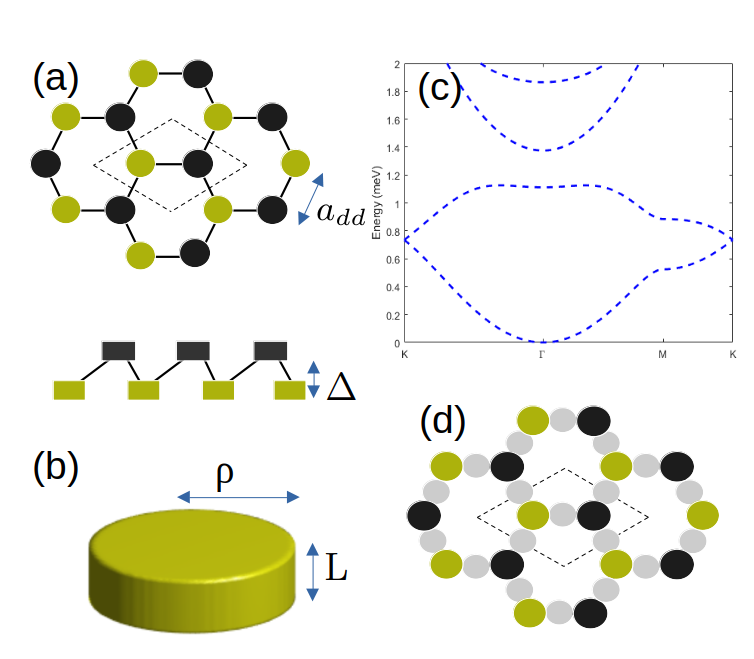}
	\caption{\label{fig1}
		(a)~Staggered honeycomb structure, with $a_{dd}$ and $\Delta$ being inter-dot distance  and relative displacement of sublattices in the out-of-plane direction, respectively. 
		(b)~Radius and length of the dots is represented by $\rho$ and $L$.
		(c)~Energy dispersion for the staggered lattice with $a_{dd}=$~50 nm, $\rho=20$~nm,  $L=10$~nm and $V_{0}=-12$~meV.
		(d)~Planar structure with repulsive potentials at the bridge sites of the honeycomb structure are shown as gray circles.\\	
	}
\end{figure}

In this work, we demonstrate the realization of both MHS and non-MHS quartic dispersions by examining the structural features of artificial semiconductor honeycomb lattices.
The rest of the paper is organized as follows. We introduce the structural features of the lattice, the theoretical aspects of quartic dispersion, the tight-binding analysis, and the numerical Bloch-wave approach in Section~\ref{sec:method}. In Section~\ref{sec:results}, we present our numerical results for the staggered and planar models. Finally, Section~\ref{sec:conclusion} summarizes our findings.

\section{Model and Methods}
\label{sec:method}

In 2D elemental crystals of group-V elements, the lattice is staggered.~\cite{sevincli_2017} Namely, the two sublattices of the honeycomb structure are displaced in opposite out-of-plane directions. Similarly, group III--VI and group IV--V honeycomb structures exhibit comparable out-of-plane displacements relative to the planar geometry.~\cite{demirci_prb_2017,ozdamar_prb_2018}
Staggering reduces the overlap between first-nearest-neighbor atoms while leaving the overlap between second-nearest neighbors essentially unchanged.
The relative strength of these overlaps plays a pivotal role in the realization of quartic dispersion, as discussed in Sec.~\ref{sec:tb}.

In order to enhance the relative strength of second--nearest--neighbor interactions in artificial lattices, we follow two approaches. In the first, we employ staggering as in real materials.
The inter-dot spacing \(a_{dd}\), dot radius \(\rho\), dot depth \(L\), sublattice shift \(\Delta\), and the applied potential \(V_0\) constitute the fundamental parameters that define the staggered structure, as shown in Fig.~\ref{fig1}(a-b).
In the second approach, a planar geometry is used.
The planar model incorporates a repulsive potential at the bridge sites in order to increase the ratio of next--nearest--neighbor to nearest--neighbor interactions, as shown in Fig.~\ref{fig1}(d).
The effect of sublattice-symmetry breaking is also investigated through an applied potential difference.

\subsection{Tight-Binding Analysis}
\label{sec:tb}

A tight-binding analysis is essential for clarifying the factors that influence the dispersion.
The Hamiltonian for a honeycomb lattice with sublattices $A$ and $B$ within the next--nearest--neighbor approximation can be written as
\begin{eqnarray}
	H = \delta V \sum_{i \in A} c_{i}^{\dagger} c_{i} 
	- t_{1} \sum_{\langle ij \rangle} 
	(c_{i}^{\dagger} c_{j}+c_{j}^{\dagger} c_{i}) \nonumber\\ 
	- t_{2} \sum_{\langle\langle ij \rangle\rangle} 
	(c_{i}^{\dagger} c_{j}+c_{j}^{\dagger} c_{i}),
\end{eqnarray}
where $i$ and $j$ are site indices, $\delta V$ is the potential difference between the sublattices, and $c_i$ ($c_i^\dagger$) annihilates (creates) an electron at the $i$-th site.
In the second and third terms, the summations run over the first and second nearest neighbors, respectively.
The corresponding solutions are given by
\begin{eqnarray} \label{eqn:TB}
	E_{\pm}=\frac{\delta V}{2}-t_2 f({\bf k}) \pm \sqrt{\frac{\delta V^2}{4}+t_1^2(3+f({\bf k}))},
\end{eqnarray}
with 
$f({\bf k})=2\cos(k_y a)+4\cos(k_y a/2)\cos(\sqrt{3}k_x a/2)$, 
$a$ is the lattice constant, and ${\bf k}$ is a vector in reciprocal space.

Both bands are filled, and the valence-band edge is centered at the center of the Brillouin zone.
Expanding $E$ for small $k$ values, one finds that for a critical value of the ratio $\xi = t_2/t_1$, the terms quadratic in $k$ vanish, giving rise to a purely quartic dispersion.
The critical value of $\xi$ is
\begin{eqnarray}
	\label{eqn:xi_c}
	\xi_c = \frac{1}{6\sqrt{1 + (\delta V / 6 t_{1})^2}},
\end{eqnarray}
and for $\xi > \xi_c$ one obtains a Mexican-hat-shaped quartic valence-band edge.~\cite{sevincli_2017}
For $\xi \le \xi_c$, the MHS feature disappears, but the quartic term remains important provided that $\xi_c - \xi$ is sufficiently small.
As a result, one can express the valence-band dispersion in a compact form as
\begin{eqnarray}
	E - E_v = -\alpha \left( k^2 + s k_{c}^{2} \right)^2.
\end{eqnarray}
Here, $\alpha$ scales the dispersion, and $k_c$ is the critical wave vector.
The type of quartic dispersion is determined by the integer parameter $s$, which can take three values:
$s = 0$ corresponds to purely quartic dispersion, while $s = -1$ and $s = 1$ correspond to MHS and non-MHS quartic dispersions, respectively.

The critical ratio given in Eq.~\ref{eqn:xi_c} is a generalization of the condition found in Ref.~\onlinecite{sevincli_2017} for nonzero $\delta V$, showing that quartic band formation is enhanced when a potential difference between the two sublattices is applied. The corresponding density of states (DOS) can be written as
\begin{eqnarray}
  g(E) &=& \frac{1}{4 \pi} \sqrt{\frac{1}{\alpha(E_v - E)}}  \nonumber \\
  & \times& \left[ 1 + \frac{s (s -1)}{2} \Theta(E - E_v + \alpha k_c^4)\right]
\end{eqnarray}
for $E<E_v$ and zero otherwise,  showing the characteristic inverse-square-root singularity at the band edge.

\subsection{Numerical Approach}

To obtain the band structure of a semiconductor artificial honeycomb lattice, we diagonalize the Schrödinger equation in momentum space following Ref.~\onlinecite{gibertini2009engineering},

\begin{equation} 
	\label{sch} 
	\det \Bigg\{ 
	\bigg[
	\frac{\hbar^2}{2m^*}(\mathbf{k} + \mathbf{G})^{2} -E_{n}(\mathbf{k})
	\bigg]
	\delta_{ \mathbf{G},\mathbf{ G }^{'} } + V(\mathbf{G} - \mathbf{ G }^{'}) 
	\Bigg\}=0,
\end{equation}
where the effective mass $m^*$ for GaAs is taken to be $0.067\,m_0$ in terms of the bare electron mass.
Here, $V(\mathbf{G})$ is the Fourier transform of the net electrostatic potential $V(\mathbf{r})$ felt by the confined electrons in the semiconductor heterostructure,
\begin{eqnarray}
	V(\mathbf{G}) = \frac{1}{\Omega}\int_{\Omega} V(\mathbf{r}) e^{-i \mathbf{G} \cdot \mathbf{r}}\, d\mathbf{r},
\end{eqnarray}
where $\Omega$ is the unit-cell volume.
We solve Eq.~\ref{sch} for both planar and staggered structures.
The reciprocal lattice vectors are generated using
$\mathbf{G} = l\, \mathbf{g}_{1} + m\, \mathbf{g}_{2} + n\, \mathbf{g}_{3}$,
where $l$, $m$, and $n$ are integers, and
{
$\mathbf{g}_{1} = {4\pi}/{\sqrt{3}a} \big[1/2, \sqrt{3}/2, 0\big]$,
$\mathbf{g}_{2} = {4\pi}/{\sqrt{3}a} \big[1/2, -\sqrt{3}/2, 0\big]$,
$\mathbf{g}_{3} = \big[0, 0, \frac{2\pi}{h}\big]$.
} 
The height of the unit-cell volume $h$ is set to 250~nm to ensure convergence and to avoid spurious interlayer effects.

In the $xy$-plane, the potential forms a sharply localized pillar due to the high-power exponential function, resembling a narrow cylindrical well.
Along the $z$-direction, each pillar is confined within a finite slab of thickness $L$, creating two disk-shaped potential wells centered at different vertical positions (e.g., $Z_1 = +10$, $Z_2 = -10$).
To model the net confinement potential $V(\mathbf{r})$, we consider a summation over cylindrical quantum-dot potentials at sites $i \in A$ or $B$, expressed in polar coordinates as
\begin{eqnarray}
	V_{i}(\mathbf{r}) &=& -V_{0i} e^{-\left(|\mathbf{r}-\mathbf{R}_{i}|^{2} / \rho^{2}\right)^\sigma} \nonumber \\ 
	&\times& \left[\Theta(z - Z_i + L/2) - \Theta(z - Z_i - L/2)\right].
\end{eqnarray}
Here, the parameter $\sigma$ controls the sharpness of the confinement potential in the $xy$-plane~\cite{19}, and we use $\sigma = 100$ in this work.
Structural parameters are chosen to be compatible with the experimental values used for realizing artificial graphene.~\cite{yue2020dirac, wang2018gaas}
For creating the antidot sites in planar geometries, a similar Gaussian model with sharpness parameter $\tilde\sigma$ is used, $\tilde V_{i}(\textbf{r}) = \tilde V_{0}\,{\exp}[-(|\textbf{r}-\textbf{R}_{i}|^{2} / \tilde \rho^{2})^{\tilde\sigma}]$ with $\tilde\sigma=2$ and $\tilde \rho=10$~nm .

\section{Results}
\label{sec:results}

\begin{figure}
\centering
\includegraphics[width=1\columnwidth]{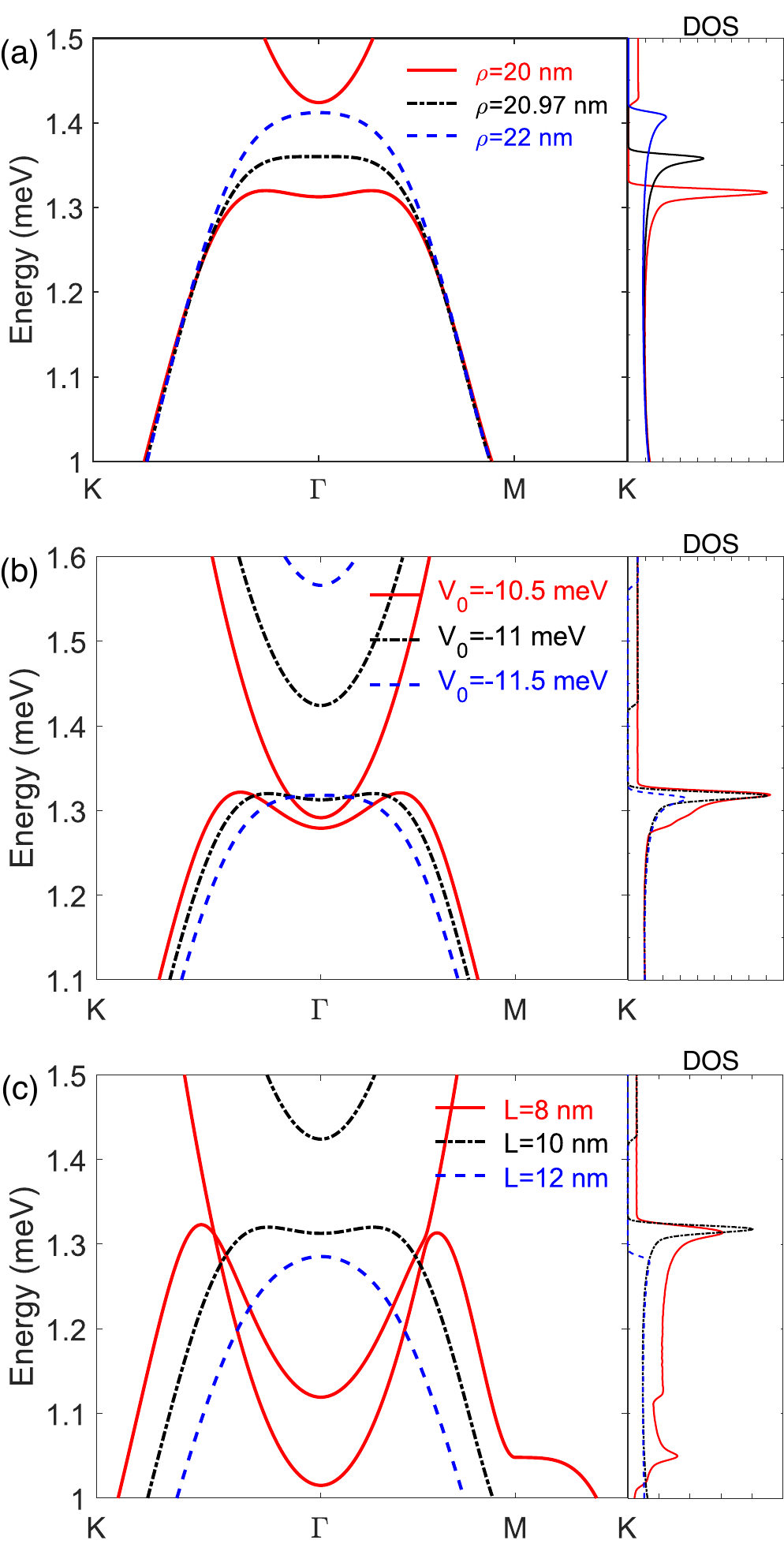}
\caption{
	Energy dispersions for various parameters,  as the dot radius is varied between 20 and 22~nm (a), the confinement voltage changes between $-10.5$ and $-11.5$~meV (b), and as the dot heights are changed from 8 to 12~nm (c).
	}
\label{Quartic} 
\end{figure}

In planar structures ($\Delta = 0$), MHS quartic dispersion is not observed over a wide range of parameters. Typically we choose $a_{dd}=50$ nm, $\rho=$ 10 to 25 nm, and $V_{0}=$ -10 to -20 meV. On the other hand, all three types of quartic dispersion can be realized in staggered ($\Delta>0$) heterostructures, depending on the structural details.
We first represent the energy dispersions of staggered configurations in Fig.~\ref{Quartic}, 
In Fig.~\ref{Quartic}(a), we show the effect of changing $\rho$ while keeping $\Delta = 20$~nm and $V_{0} = -11$~meV.
We note that increasing $\rho$ mainly enhances the second--nearest--neighbor hopping $t_{2}$.
For the smallest value, $\rho = 20$~nm, a clear MHS is observed.
At $\rho = 20.97$~nm, a purely quartic dispersion ($s = 0$) is realized.
As $\rho$ is increased further to 22~nm, the band has a non-MHS quartic character ($s=1$), whose signature is the enhanced DOS at the band edge.

The effect of changing the confinement potential $V_{0}$ is shown in Fig.~\ref{Quartic}(b), where the vertical shifting and dot-radius parameters are fixed at $\Delta = 20$~nm and $\rho = 20$~nm, respectively.
For $V_{0} = -10.5$~meV, although a MHS quartic dispersion exists, a band gap has not yet formed.
Decreasing $V_0$ causes the orbitals to become more localized at their sites, diminishing both $t_1$ and $t_{2}$ in a nontrivial manner.
For $V_{0} = -11$~meV, a band gap opens accompanied by a MHS dispersion.
For $V_{0} = -11.5$~meV, the band gap increases to 3~meV while the dispersion becomes non-MHS, a behavior similar to that observed when $\rho$ is increased.

In Fig.~\ref{Quartic}(c), $\Delta = 15$~nm, $\rho = 20$~nm, and $V_0 = -11$~meV are fixed, while the dot heights are varied from 8 to 12~nm.
MHS quartic valence bands are realized for both $L = 10$~nm and $L = 12$~nm.
However, higher-energy bands overlap for $\Delta = 12$~nm, leading to metallic behavior.
For larger values of $\Delta$, a gap opens and increases with $\Delta$, while preserving the MHS dispersion.
The quartic-dispersion behavior is also clearly reflected in the DOS, showing a peak at the band edge that becomes sharper with increasing $\Delta$.
We note that when $\Delta$ becomes too large, i.e., $t_{2} \gg t_{1}$, the system behaves like two weakly coupled triangular lattices (not shown).

\begin{figure}
	\centering
	\includegraphics[width=0.48\textwidth]{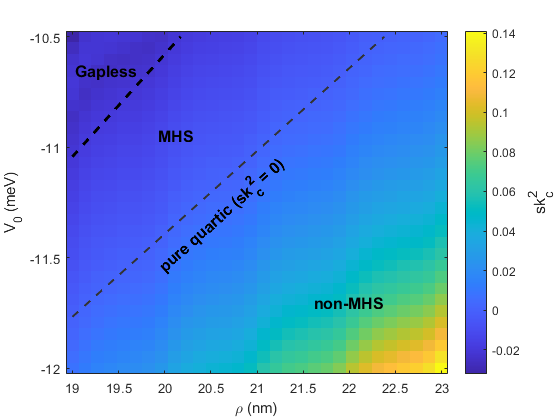}
	\caption{
		Color map of the extracted coefficient $s\,k_{c}^{2}$ as a function of $\rho$ and potential depth $V_{0}$. Negative values ($s\,k_{c}^{2} < 0$) indicate Mexican-hat-shaped dispersion with a ring-like minimum, while positive values correspond to conventional parabolic bands centered at $\Gamma$. The dashed line marks the $s\,k_{c}^{2} = 0$ condition, indicating purely quartic dispersion. The diagram highlights three regimes: gapless (band touching), MHS (quartic with indirect gap), and non-MHS (parabolic). This phase map guides the design of quartic band structures.\\
	}
	
	\label{fig:etakc2_phase}
\end{figure}

Fig.~\ref{fig:etakc2_phase} presents a color map of the extracted coefficient $s k_{c}^{2}$ as a function of the potential depth $V_{0}$ and dot spacing $\rho$. 
The coefficient $s k_{c}^{2}$ characterizes the curvature of the conduction band near the $\Gamma$ point: negative values indicate a MHS quartic dispersion. The dashed line marks the condition $s k_{c}^{2} = 0$, which defines a purely quartic dispersion.
Based on the band topology and gap structure, the parameter space is classified into three regions: a \textit{gapless region}, where the second and third bands overlap; an \textit{MHS region}, with a finite indirect band gap and quartic dispersion; and a \textit{non-MHS quartic region}, where the maximum remains centered at $\Gamma$ without an MHS, but the peak in the DOS at the band edge persists as the signature of a quartic dispersion relation. 
This diagram provides a systematic framework for identifying regimes that support quartic band structures through geometric and potential engineering.

\begin{figure}
	\centering
	\includegraphics[width=1\columnwidth]{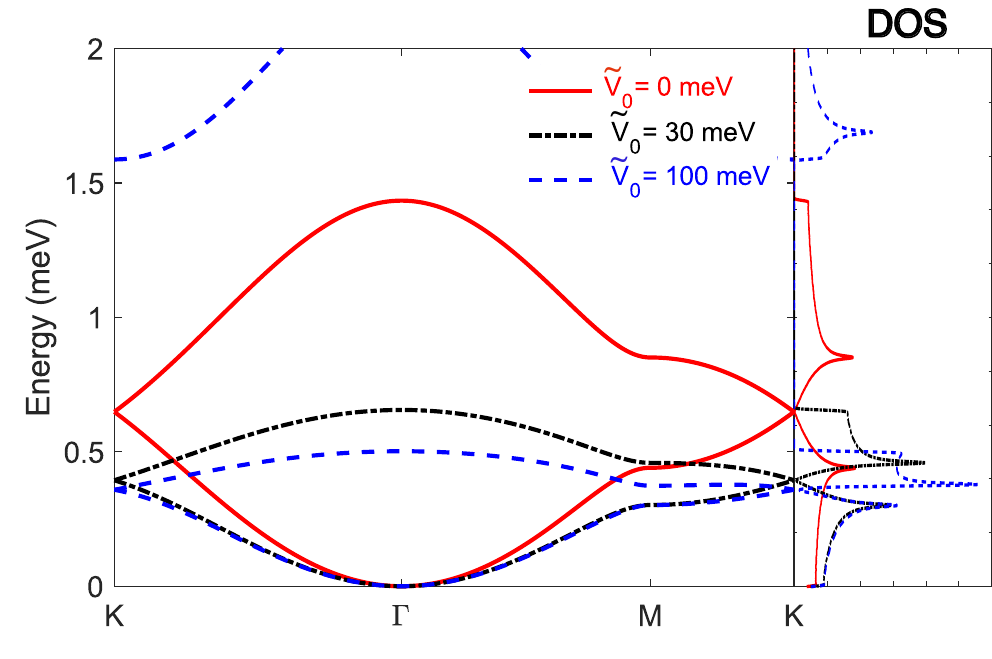}
	\caption{\label{fig:planar} 
		Solid lines represent the energies of classical two-dot unit cells. Attractive potentials 70 meV and 100 meV energies demonstrated by dotted-dashed lines.
	}
\end{figure}

\begin{figure}
	\centering
	\includegraphics[width=1.\columnwidth]{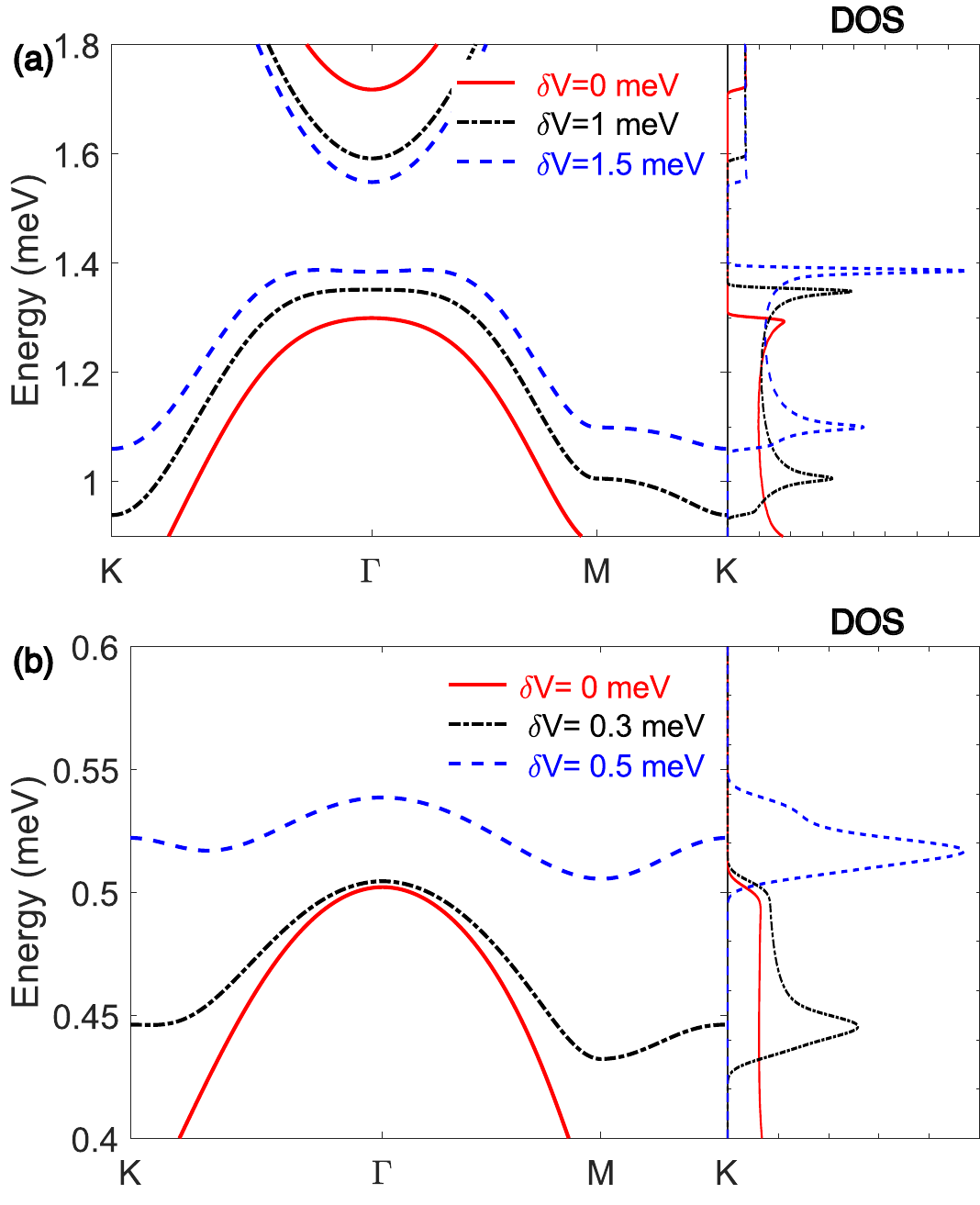}
	\caption{\label{fig:deltaV} 
		Effects of sublattice potential differences as calculated for the  {staggered} (a), and planar (b) geometries.
	}
\end{figure}

Next, we investigate the planar model, which is schematically shown in Fig.~\ref{fig1}d.
Instead of vertically displacing the A and B sublattices, a repulsive potential with strength $\tilde V_{0}$ is inserted between every nearest neighbor in order to increase the $t_2/t_1$ ratio.
The resulting energy dispersions are shown in Fig.~\ref{fig:planar} for $\rho = 10$~nm.
For $\tilde V_{0} = 0$, a standard honeycomb-lattice dispersion is reproduced, as expected.
For $\tilde V_{0} = 30$~meV, an overall squeezing of all bands is observed as a result of reduced $t_1$.
However, no MHS dispersion forms even at $\tilde V_{0} = 100$~meV.
A signature of non-MHS quartic dispersion can be observed in the DOS as a peak forming near the band edge, showing that although a quartic energy--momentum relation is possible, the formation of an MHS dispersion is not feasible in a planar lattice.

Lastly, we focus on the effect of a potential difference between the sublattices, $\delta V = V_{0A} - V_{0B}$.
As examined within the tight-binding approximation (Eq.~\ref{eqn:TB}), $\delta V$ should enhance the formation of quartic band dispersion in both planar and staggered geometries.
In Fig.~\ref{fig:deltaV}a, we show the dispersions for the staggered geometry, where $\rho = 20$~nm, $\Delta = 20$~nm, and $V_{0A} = -12$~meV.
For $\delta V = 0$, this choice of parameters leads to a relatively weaker singularity.
However, as $\delta V$ is increased to 1.5~meV, an MHS dispersion forms, and a sharp DOS peak appears, confirming the tight-binding analysis.
On the other hand, for the planar model with $\rho = 10$~nm, $\Delta = 20$~nm, and $V_{0A} = -10$~meV, the valence band becomes severely distorted for values of $\delta V$ as small as 0.5~meV, but no signature of quartic dispersion appears at the band edge.

\section{Conclusion}
\label{sec:conclusion}

In summary, we have shown that artificial semiconductor honeycomb lattices enable the controlled realization of MHS, purely quartic, and non-MHS quartic dispersions. Analytical tight-binding theory and numerical Bloch-wave calculations together identify the key parameters governing quartic-band formation, including dot radius, confinement depth, vertical staggering, and sublattice potential asymmetry. Staggered geometries reliably produce MHS-type dispersions through enhanced second--nearest--neighbor coupling, whereas planar lattices yield only purely quartic or non-MHS quartic bands.
The resulting phase diagram identifies the conditions for distinct classes of quartic dispersion. Artificial graphene thus provides a practical pathway for engineering quartic band topologies relevant to correlated quantum phases.

\begin{acknowledgments}
  This work was supported by the Izmir Institute of Technology Scientific Research Projects under the program of ADEP (Project No: 2022IYTE-3-0031). 
  HS acknowledges   support from Air Force Office of Scientific Research (AFOSR, FA9550-21-1-0261)
\end{acknowledgments}

\bibliography{main}

\end{document}